\documentclass[a4paper,twocolumn,11pt,accepted=2024-05-15]{quantumarticle}
\pdfoutput=1

\usepackage{graphicx}
\usepackage{dcolumn}
\usepackage{bm}
\usepackage{hyperref}
\usepackage{fancyhdr}
\usepackage{todonotes}
\usepackage{tikz}
\usetikzlibrary{quantikz}

\newcommand{\code}[1]{\texttt{#1}}

\begin{document}

\title{A High Performance Compiler for Very Large Scale Surface Code Computations}

\author{George Watkins}
\email{invio.george@gmail.com}
\affiliation{Department of Computer Science, Aalto University, 00076 Espoo, Finland}
\affiliation{School of Computing Science, Simon Fraser University, Burnaby, B.C., Canada V5A 1S6}

\author{Hoang Minh Nguyen}
\email{hoangminh98@gmail.com}
\affiliation{School of Computing Science, Simon Fraser University, Burnaby, B.C., Canada V5A 1S6}

\author{Keelan Watkins}
\email{keelan\_w@outlook.com}
\affiliation{Department of Physics, Simon Fraser University, Burnaby, B.C., Canada V5A 1S6}

\author{Steven Pearce}
\email{stevenp@sfu.ca}
\affiliation{School of Computing Science, Simon Fraser University, Burnaby, B.C., Canada V5A 1S6}

\author{Hoi-Kwan Lau}
\email{hklau.physics@gmail.com}
\affiliation{Department of Physics, Simon Fraser University, Burnaby, B.C., Canada V5A 1S6}
\affiliation{Quantum Algorithms Institute, Surrey, B.C., Canada V3T 5X3}

\author{Alexandru Paler}
\email{alexandrupaler@gmail.com}
\affiliation{Department of Computer Science, Aalto University, 00076 Espoo, Finland}

\begin{abstract}
We present the first high performance compiler for very large scale quantum error correction: it translates an arbitrary quantum circuit to surface code operations based on lattice surgery. Our compiler offers an end to end error correction workflow implemented by a pluggable architecture centered around an intermediate representation of lattice surgery instructions. Moreover, the compiler supports customizable circuit layouts, can be used for quantum benchmarking and includes a quantum resource estimator. The compiler can process millions of gates using a streaming pipeline at a speed geared towards real-time operation of a physical device. We compiled within seconds 80 million logical surface code instructions, corresponding to a high precision Clifford+T implementation of the 128-qubit Quantum Fourier Transform (QFT). Our code is open-sourced at \url{https://github.com/latticesurgery-com}.
\end{abstract}

\maketitle

\section{Introduction}

Applying surface quantum error correcting codes (QECCs) efficiently to large computations is challenging in terms of classical computing resources necessary for the compilation process. Compilers tailored for QECC are only starting to appear, often with significant limitations with respect to the scale of the circuits that can be handled or the compilation time.

Large scale QECC compilation is a necessity, because practical algorithms, like Shor's and Grover's assume high-quality qubits with a very low error rate~\cite{suchara2013comparing}, but we are unlikely to obtain hardware (physical) qubits with such fidelity in the near future~\cite{preskill2018quantumcomputingin}. QECCs solve this issue by using a large number of error prone physical qubits to encode higher fidelity \textit{logical} qubits. For example, a quantum factoring algorithm needs roughly $1000$ qubits to factor a $1000$-bit number and millions of gates~\cite{parker2000efficient, gidney2021factor}. Consequently, practical algorithm require very large scale quantum computers, while only some carefully crafted examples of problems where quantum hardware has an advantage with small devices exist~\cite{arute2019quantum}. 

Surface codes are a family of QECCs that require low qubit connectivity and a reasonably high hardware error rate (such as between $0.1\%$ and $1\%$) to create good logical (computational) qubits~\cite{fowler2012surface, wang2011surface, tomita2014low} and only require degree four nearest neighbour connectivity. These properties make them a promising option for error correcting devices with a couple hundred logical qubits. Physical devices with compatible layouts have already been made or proposed, albeit on a small scale~\cite{arute2019quantum, andersen2020correction,andersen2020detection,acharya2022suppressing, bourassa2021blueprint}. Examples of larger scale quantum circuits protected by surface QECCs were compiled manually in \cite{hanks2020effective, gidney2019flexible}. The complexity of optimising surface code circuits has been shown to be related to NP-hardness~\cite{herr2017optimization, wasa2023hardness}.

\begin{figure*}[!t]
    \centering
    \includegraphics[width=1.9125\columnwidth]{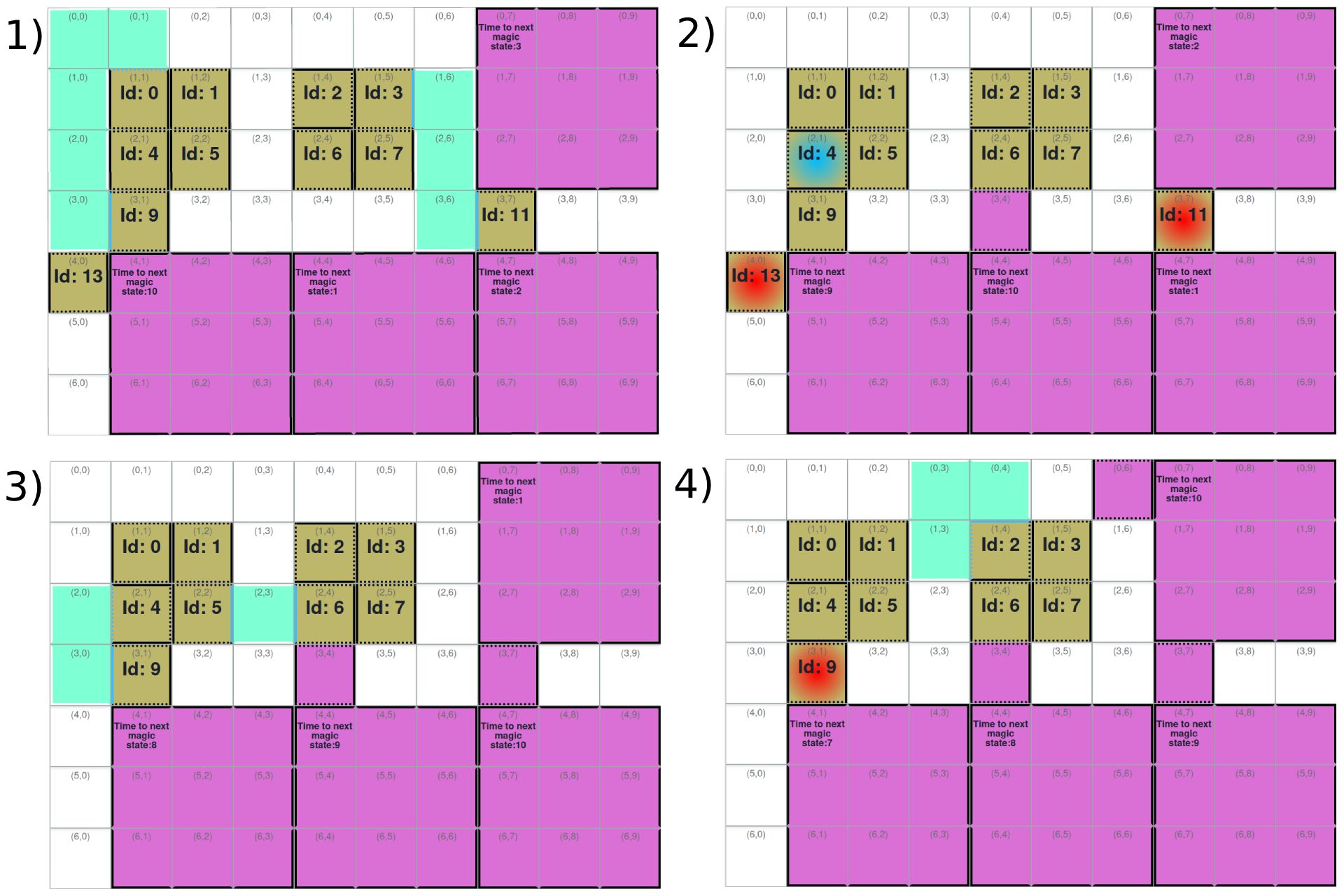}
    \caption{Example output of from the  compiler: sequence of discrete time steps, called \emph{slices}, of a surface code computation. Each slice has an associated time instant when it is taking place. The slices are obtained after mapping a circuit to a layout, where patches end up being used for holding error-corrected logical qubits (brown), distillation procedures (pink), routing merge and split operations of patches (blue), or not used (white). 1) There is a merge and split operation (blue) taking place between logical qubits 0 and 3. 2) Qubits 11 and 13 are measured; 3) logical operation between qubits 4 and 9, and logical operation between 5 and 6, and the bottom right distillation is outputing a distiiled state; 4) qubit 9 is measured, merge and split between 1 and 2, and the right most distillation region is outputing a distilled state. 
    }
    \label{fig:LargeScaleComputation}
\end{figure*}

We present and demonstrate the extremely high scalability of our efficient QECC compiler. This is a step forward for quantum software: we create a streaming pipeline and a compilation environment for the compilation and optimisation of very large scale QECCs. Our high performance pipeline makes it possible to process extremely large circuits (would not fit in memory). We can compile directly, in a streaming process, by reading and writing to mass storage. Streaming enables the real-time operation of our compiler, meaning that this tool may be integrated in the classical control software necessary to operate quantum computers~\cite{paler2020aggregated}.

This paper is organised as follows: In Sec.~\ref{sec:back} we introduce the concepts necessary for presenting the compilation methods and workflow. Sec.~\ref{sec:methods} describes the two-stage compilation pipeline that consists of gate level processing (Sec.~\ref{sec:TransformingTheGateSet}) and logical operation routing (Sec.~\ref{sec:routing}). The latter includes also a fast method to perform state vector simulation that takes into account the entangling and disentangling action of the lattice surgery operations. Finally, Sec.~\ref{sec:res} illustrates the performance of our compiler. We compile within seconds a high-precision 128-qubit Quantum Fourier Transform (QFT)~\cite{nielsen2001quantum}. To the best of our knowledge, this is the largest-scale compilation of this kind.

\section{Background}
\label{sec:back}

This section introduces the necessary background details for describing the compilation process.
The application of error correction to quantum circuits resembles the process well known to classical computer scientists of program compilation: the \textit{compiler} reads code in a programming language (higher level quantum gates) and outputs machine instructions (lattice surgery quantum gates).

We opted for flexibility and developed a compiler with a well-defined intermediate representation to separate circuit pre-processing from surface code instruction layout. Surface code instructions for large scale computations at present is interesting for at least two purposes, one is being able to produce reliable resource estimates, and the other is to start preparing for when we will have such devices, so that hardware engineers can start designing devices with instruction sets for error correction in mind.

We assume the reader is familiar with the basic concepts of quantum computing and quantum information~\cite{nielsen2001quantum, mermin2007quantum}. We assume the conventional meaning for common quantum gates (Phase gates S and T, Hadamard gate H, CNOT, Toffoli) and the Pauli matrices (I, X, Y, Z). By the phase rotation gate $R_Z(\theta)$ we mean:
\begin{equation*}
    R_Z(\theta) = 
    \begin{bmatrix}
    1 & 0 \\
    0 & e^{i\frac{\theta}{2}}
    \end{bmatrix}
\end{equation*}

We will frequently use Pauli product rotations, for which we assume the following: given an axis P (which may be a Pauli matrix or a tensor product of Pauli matrices) we denote by $P(\theta) = \text{exp}( - i\theta P) = \cos(\theta)I - i \sin(\theta)P$. Note that under this convention, $R_P(\theta) = P(\frac{\theta}{2})$ for $P=X,Z$. Also, when the Pauli matrices appear with sub-indices, e.g. $Z_1Z_2Z_3Z_4$ in Fig.~\ref{fig:Stabilizers}, we mean the tensor product $X \otimes Z$ of the Pauli matrices applied to qubits indexed $1$ and $2$. Similarly, we use gate $R_X(\theta)=HR_Z(\theta)H$.

\subsection{Surface Codes}
\label{Sec:IntroLatticeSurgery}

\begin{figure}[!t]
    \centering
    \includegraphics[width=0.75\columnwidth]{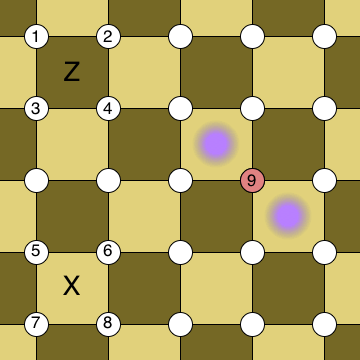}
    \caption{A graphical depiction of surface code layout. The white circles are \textit{data} qubits protected from errors by measuring stabilizers around them. The squares in the lighter and darker shades of yellow represent stabilizer measurements. For example, the squares marked with $Z$ and $X$ represent the $Z_1 Z_2 Z_3 Z_4$ and $X_5 X_6 X_7 X_8$ stabilizer measurements respectively. If an error occurs in a data qubit, such as a phase flip occurring on 9, the $X$ stabilizers around it will pick it up by changing outcome (\textit{syndromes}, highlighted in purple). There are advanced methods to decode sets of errors (e.g~\cite{fowler2012surface,varsamopoulos2020decoding,hu2020quasilinear}). Errors can either be corrected on the spot or tracked classically by inverting later readouts. This cycle of detecting, decoding and correcting is referred to as the \textit{surface code cycle}.}
    \label{fig:Stabilizers}
\end{figure}

A major challenge with the current generation of quantum computers is the occurrence of errors while performing computations. Errors may occur because of control system faults or stray interaction with the environment. A proposed solution for avoiding errors are Quantum Error Correcting Codes (QECC). These codes add some degree of fault tolerance to computations by using many \textit{physical qubits} to form fewer but more reliable abstract \textit{logical qubits}\cite{terhal2015quantum}. Surface codes are a family of QECCs that aim at improving computational fidelity by entangling physical qubits in a \emph{physical lattice}~\cite{roffe2019quantum, nielsen2001quantum}. This kind of codes, with topological properties, was first theorized with exotic particles known as ``anyons''~\cite{kitaev2003fault}. Surface codes are appealing because they are well understood, and feature a high error threshold. In near future, quantum computing hardware with thousands of qubits might be realized~\cite{googleResearchJourney,IBMResearchRoadmap,bourassa2021blueprint} which would be able to
 operate a surface code cycle on a lattice of qubits.

The key step of surface code error detection is stabilizer measurement, as shown by the shaded squares in Fig.~\ref{fig:Stabilizers}. These measurements act as parity checks on bit flips or phase flips of a square lattice of \textit{data qubits}. The surface code and its cycle (the sequence of quantum gates applied for enforcing the code constraints) only tell us how to protect a lattice from error. The surface code distance indicates how much error is tolerated~\cite{bravyi1998quantum}.

\begin{figure}[!t]
    \centering
    \includegraphics[width=\columnwidth]{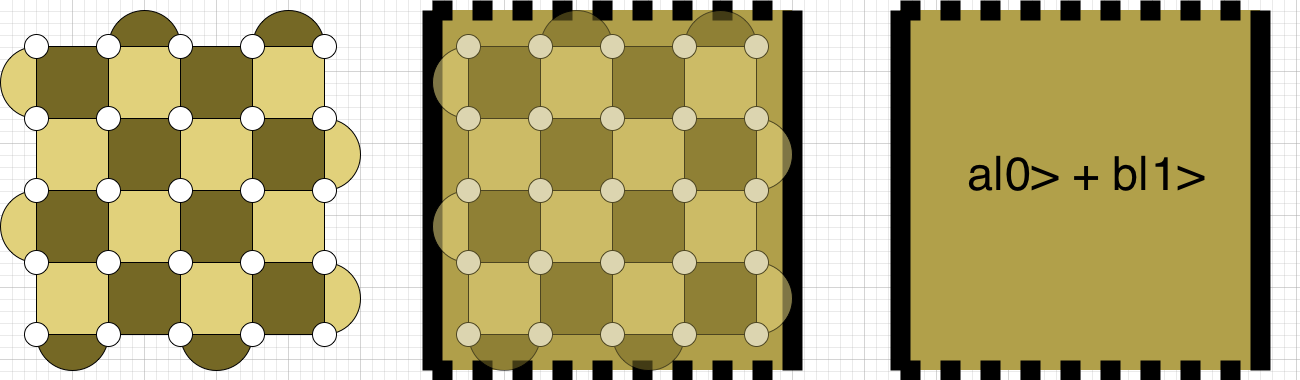}
    \caption{Abstracting physical qubits to patches. We omit the details of stabilizers and data qubits that make up patches, and instead represent distance-independent features. It is always possible to compute back these details about stabilizers from the output format and code distance. The picture to the left shows how this abstract representation relates to the physical implementation, and to the right there is a fully abstract patch, which has its own logical state. The different stabilizers on the boundaries yield two different kinds of boundaries, which are often referred to as \textit{rough} and \textit{smooth}.}
    \label{fig:singlepatches}
\end{figure}

\subsection{Logical Qubits and Logical Operations}
\label{sec:LogicalOps}

Logical (computational) qubits are encoded by ``cutting out'' portions of a device's physical lattice into \textit{patches}, which are cluster states error corrected by the surface code cycle. This encoding of logical qubits is known as the \textit{planar code}~\cite{bravyi1998quantum,dennis2001topological}. Patches have boundaries outside of which they don't interact, except when performing certain logical operations (Sec.~\ref{sec:LogicalOps}). Fig.~\ref{fig:singlepatches} outlines how patches relate to the surface code. The patch-based approach has been shown to be a resource-efficient choice for quantum error correction~\cite{horsman2012surface, poulsen2017fault, litinski2019game}.

\begin{figure}[t]
    \centering
    \includegraphics[width=\columnwidth]{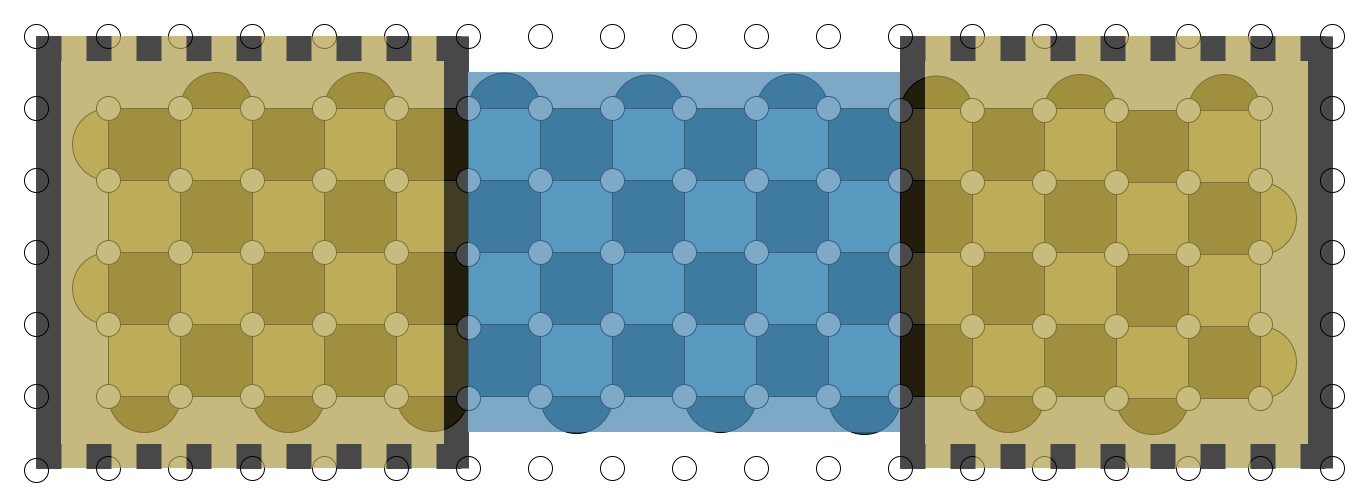}
    \includegraphics[width=\columnwidth]{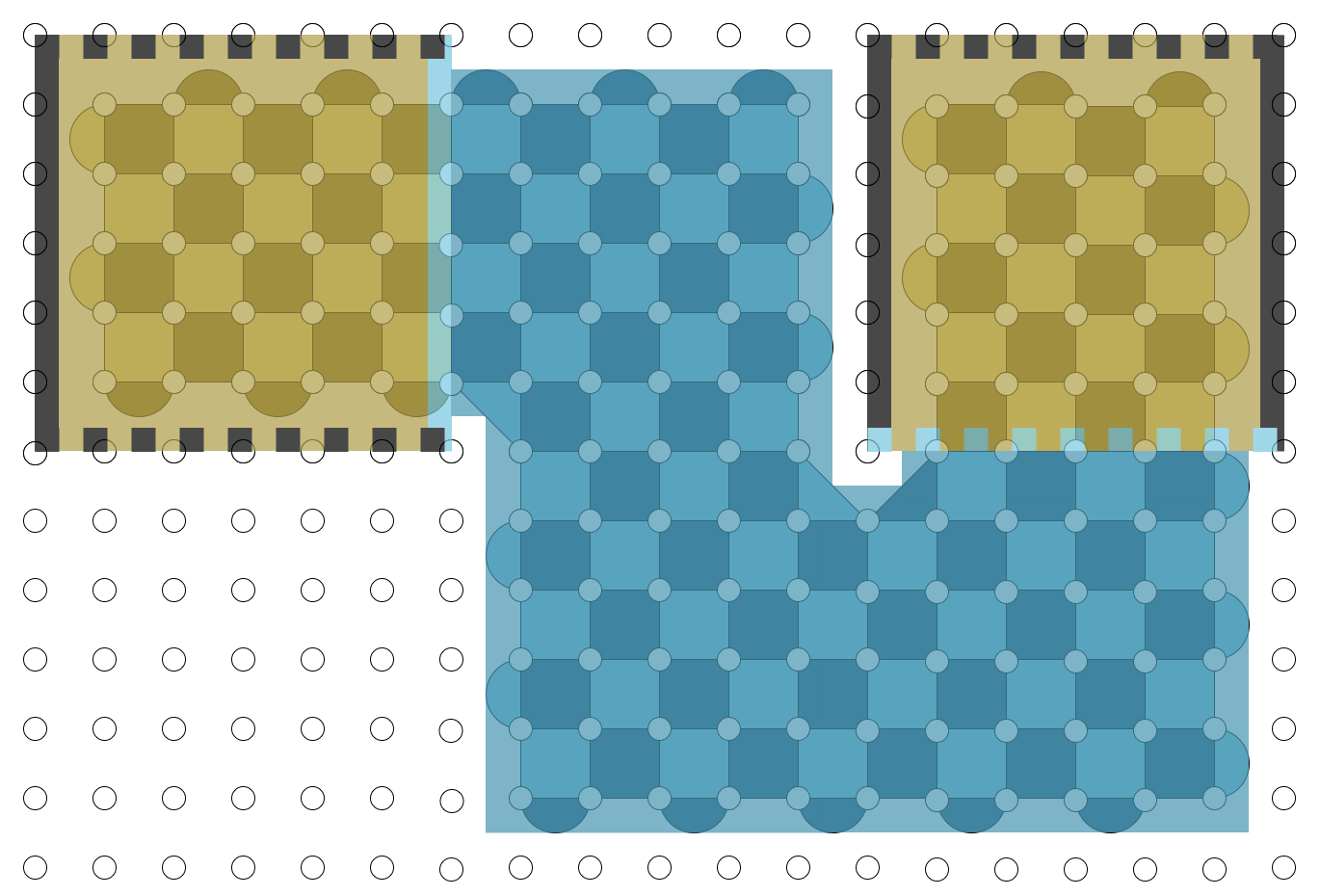}
    \caption{Lattice surgery of patches. Patches are \textit{merged} by activating the stabilizer measurements with the data qubits between them (blue regions). This operation causes the two patches to become one, hence losing a degree of freedom and projecting the logical state into a subspace. After stopping the stabilizer measurements and measuring the mediating data qubits, the patches are \textit{split}. Overall, this operation is equivalent to a logical multi body measurement~\cite{horsman2012surface}. The observable depends on the boundaries: rough for $X$ and smooth for $Z$. This figure shows measurements of the observables $Z\otimes Z$ (top) and $Z\otimes X$  (bottom)}
    \label{fig:mergeSplit}
\end{figure}

We will be looking at square patches with two kinds of boundaries that encode a single qubit. Patch size is proportional to code distance and the performance of the decoding algorithm (e.g.~\cite{fowler2012surface, hu2020quasilinear}). For our intents, it suffices to know that the size of the patches will depend on the physical error rate of the device, length of the computation and desired success rate of the logical computation. In Sec.~\ref{sec:res} we estimate the resources~\cite{herr2019time} necessary to execute the compiled output.

Having obtained logical qubits, we require a method to perform operations between them. Table~\ref{table:SurfaceCodeOp} offers an overview of all the surface code operations supported by our compiler at the logical level. Some logical operations are performed directly on patches: Pauli X and Z~\cite{bravyi2005universal}, and Hadamard gates~\cite{horsman2012surface}, can be implemented in this way and are called \textit{transversal} operations. It is also possible to directly initialize a patch in the $\ket{0}$ or $\ket{+}$ states and to measure in the X or Z basis~\cite{litinski2019game}. For the remaining operations needed to complete a universal gate set we use lattice surgery~\cite{horsman2012surface}. This protocol achieves entangling multibody measurements by merging and splitting patches. 

We use these measurements along with prepared ancillae states (and corresponding patches) to implement CNOT as shown in~\cite{horsman2012surface}, and the S and the T gates (Fig.~\ref{fig:GatesToLLI} in Appendix). T gates utilize a \textit{magic state}, $\ket{m} = \frac{1}{\sqrt{2}}(\ket{0}+e^{\frac{i \pi}{4}}\ket{1})$, which in the surface code cannot be initialized directly with a high fidelity. These states have to be prepared by \textit{distillation}. There are several protocols for magic state distillation~\cite{bravyi2012magic,litinski2019magic}, but for our compilation purposes it suffices to acknowledge the fact that these distillations occupy some amount of space on the device's lattice and that they have a certain duration in time: distillation regions are described by their \emph{bounding box} which includes a time axis for how long it will take to produce the next magic state.

\subsection{Related Work}

Compilers for surface codes have been previously presented in the literature, and most of the times, the compilation problem has been decoupled from the challenges of optimising the resulting circuits. In general, automatic optimisation is performed by implementing heuristic algorithms for the efficient layout of the logical operations, and this includes parallelizing as many as possible operations, using fewer patches for the routing etc. Surface code computations can be implemented through braiding (e.g. Surfbraid~\cite{paler2019surfbraid}) or lattice surgery like the tool presented herein (e.g. OpenSurgery~\cite{paler2019opensurgery} or the compilers from~\cite{lao2018mapping, hua2021autobraid,beverland2021surface}).

Our compiler is distinct from the others in the following ways. Our compiler's source and target are similar to OpenSurgery, but improves on the compilation time performance, offers new optimizations, adds the ability to customize layouts, and handles parallel magic state distillation.

The compilers from~\cite{hua2021autobraid,beverland2021surface} focus specifically on routing long range surface code interactions. While we do tackle such problem, as it is necessary for our overall compilation goal, the focus of this project is broader in scope and we organise our compilation into a very modular, highly efficient pipeline which can handle both short and long range interactions.

Our compiler supports very large scale layouts through a layout specification and the compiler can automatically map large-scale circuits to the layouts. In contrast, the compiler from \cite{lao2018mapping} is a small scale procedure that explores the trade offs of different layouts for mapping algorithms onto surface code architectures.

Our compiler is modular and can include manual optimisation techniques, for example, by replacing existing gate decompositions, or reconfiguring the bounding boxes of the distillation sub-circuits. For completeness, manually obtained surface code layouts with techniques such as the AutoCCZ for optimizing ripple carry adders where presented for example by~\cite{gidney2019flexible}. Finally, one last approach to quantum compilers worth mentioning are variational compilers~\cite{xu2019variational}, which share with our project the challenges of circuit pre-processing.

Compared to existing surface code compilers, our tool extends the state of the art by including at least the following novelties:
\begin{itemize}
    \item support for an intermediate language for compiling high-level circuits from different languages (e.g. Q\#, Cirq, Qiskit) and descriptions (e.g. Clifford+T, multibody measurements);
    \item highly configurable layouts for qubits, routing space and multiple parallel distillation procedures;
    \item very high-speed, configurable routing heuristics which can be easily replaced with more sophisticated approaches including based on machine learning;
    \item pipelined, modular design that is compatible with distributed computing platforms such that compilations and optimisations can be performed on multi-core/paralell computers.
\end{itemize}

\begin{table}[t]
  \centering
  \begin{tabular}{|p{.35\columnwidth}| p{.55\columnwidth}|}
    \hline
        \textbf{Operation} & \textbf{Method}\\
    \hline
    \hline
        Patch initialization in the $\ket{0}$ and $\ket{+}$ states & Direct Initialization of data qubits~\cite{horsman2012surface} \\
    \hline
        Single patch measurements
        & Direct measurement of data qubits~\cite{horsman2012surface} \\
    \hline 
        Pauli X and Z & Transversal in surface codes~\cite{fowler2012surface} \\
    \hline
        Hadamard gates
        & Transversal in planar code patches~\cite{horsman2012surface} \\
    \hline
        Entangling multi body measurements
        & Lattice surgery merges and splits, mediated by ancilla patches for routing~\cite{litinski2019game} \\
    \hline 
        Boundary Rotation
        & Patch deformation\cite{litinski2019game} \\
    \hline 
        S gates &
        Lattice surgery with twist defects~\cite{litinski2018lattice} \\
    \hline 
    Preparation of Magic states $\ket{m} = \ket{0}+e^{\frac{i \pi}{4}}\ket{1}$
        & Distillation in dedicated regions~\cite{litinski2019magic} \\
    \hline
  \end{tabular}
  \caption{The list of logical surface code operations supported by the compiler. The operations are formalized into \textit{logical lattice instructions} (LLI), which serve as a central intermediate representation to our compiler. LLI decouples the pre-processing to surface code instructions from laying them out on an abstract lattice.}
  \label{table:SurfaceCodeOp}
\end{table}

\section{Methods}
\label{sec:methods}

We address the problem of taking a circuit specified in a machine readable format, and converting the circuit to the surface code operations outlined in Table~\ref{table:SurfaceCodeOp}. For small circuits it is easy enough to perform such conversion by hand, but automation is necessary for large scale circuits.

Our \textit{compiler} is a computer program that reads text in a \textit{source} formal language and outputs machine code in another language, called the \textit{target}. In our case the source is a quantum circuit in a subset of OpenQASM 2.0~\cite{cross2017open} (Sec.~\ref{sec:qasmmin}), while the target is a JSON logical operation instructions (Sec.~\ref{sec:LogicalOps} and Fig.~\ref{fig:LargeScaleComputation}).

We implemented our compiler and the source code is open sourced at \url{https://github.com/latticesurgery-com}. In order to improve the readability of this paper, we keep the engineering and implementation details to a minimum, and point the interested reader to the open sourced code. The latter is written with modern C++ features which increase the comprehension of the code's functionality.

The compiler is continuously tested and verified for functional correctness with modern continuous integration, while practical performance plays a significant role. Our compiler offers a wide range of configuration options, ranging from optimization heuristics, intermediate representations of the computations, as well as flexible layouts.

\subsection{The Compilation Pipeline}

\begin{figure}[t]
    \centering
    \includegraphics[width=\columnwidth]{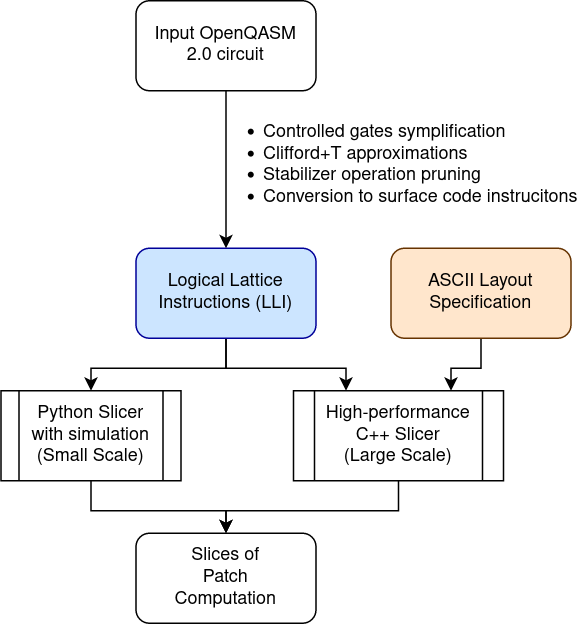}
    \caption{The pipeline as implemented in the compiler.}
    \label{fig:pipeline}
\end{figure}

The compiler operates a two-stage pipeline (Fig.~\ref{fig:pipeline}): 1) a pre-processing stage, and 2) a layout and routing stage. The two stages communicate through an intermediate representation we refer to as \emph{logical lattice instructions} (LLI from Table~\ref{table:SurfaceCodeOp}). The LLI contains all the information about the logical operations happening on the lattice, but none about the physical locations of the patches, or about routing and distillation regions. The physical qubit lattice will be operated according to LLI instructions (Table~\ref{table:SurfaceCodeOp}). 

The first stage, the \emph{gate level processing stage}, operates mostly at the logical circuit level. We resort to a universal gate set based on surface code operations. We gradually process the input circuit's gates to align with our surface code instructions. Once the circuit is in a suitable format (only Clifford+T gates or certain Pauli rotations), the circuit maps 1-to-1 with surface code operations and is written down as LLI.

The second stage is the \textit{slicer}. Herein, the LLI are combined with a layout specification in space and time (Fig.~\ref{fig:layout}). The LLI language is circuit layout agnostic, meaning that the mapping of the logical qubits to the physical lattice may have a great impact on the efficiency of the compiled circuit. The result of these steps is a ``sequence" of \textit{slices} of the physical lattice. The slices depict the state of the computation at each point in time, as shown in (Fig.~\ref{fig:LargeScaleComputation}). We offer two such slicers: one written in Python, geared towards the verification of small scale circuits (Sec.~\ref{sec:slowSlicer}) and a high performance one written in C++ for large scale circuits (Sec.~\ref{sec:cppSlicer}).

\subsection{Gate Level Processing}
\label{sec:TransformingTheGateSet}

The first stage takes a logical circuit specified in our own minimal dialect of OpenQASM 2.0. We offer two ways to pre-process the circuit: 1) with Pauli rotations and Pauli product measurements, and 2) directly with higher level quantum gates such as Toffoli gates. In both cases, we first parse the circuit into a list of gates, using either Qiskit~\cite{Qiskit}, PyZX~\cite{kissinger2020Pyzx}, a custom parser or a combination of the three depending on the circuit.

The gate list expression of the input circuit might use gates which are not supported by the error-correction procedure. In this step we reduce the gate set so that it easily translates to LLI. Our custom parser is able to break down very small angle rotations, such as $Z(\frac{\pi}{2^{128}})$ by symbolic processing of the argument. These rotations are needed to compile, for example, a 128-qubit quantum Fourier transformation (QFT) circuit. After parsing, the list of gates is passed through the pipeline to the next stage.

First, controlled gates are broken down to CNOTs and single qubit rotations using the identity in Fig.~\ref{fig:crz}. The circuit now only has single qubit Clifford gates, CNOTs and single qubit rotations. At the last stage of pre-processing in the gate model single qubit rotations smaller than $\frac{\pi}{4}$ are approximated to single qubit Clifford+T gates.

It is possible to convert controlled-rotation gates to Clifford operations plus some small angle $Z(\theta)$ rotations (Figure~\ref{fig:crz} in Appendix). The latter are not Clifford+T and are difficult to perform in a fault-tolerant way~\cite{jochym2018disjointness}. We achieve arbitrary $Z(\theta)$ rotations by approximating them with Clifford+T gates, for which we leverage the Gridsynth package~\cite{ross2014optimal} which outputs approximations constituted of sequences of H, X, Z, S and T gates. The T gates are performed by consuming magic states, which are prepared in dedicated \textit{distillation regions}\cite{bravyi2005universal,litinski2019magic}.
  
\begin{figure}[t]
    \centering
    \includegraphics[width= \columnwidth]{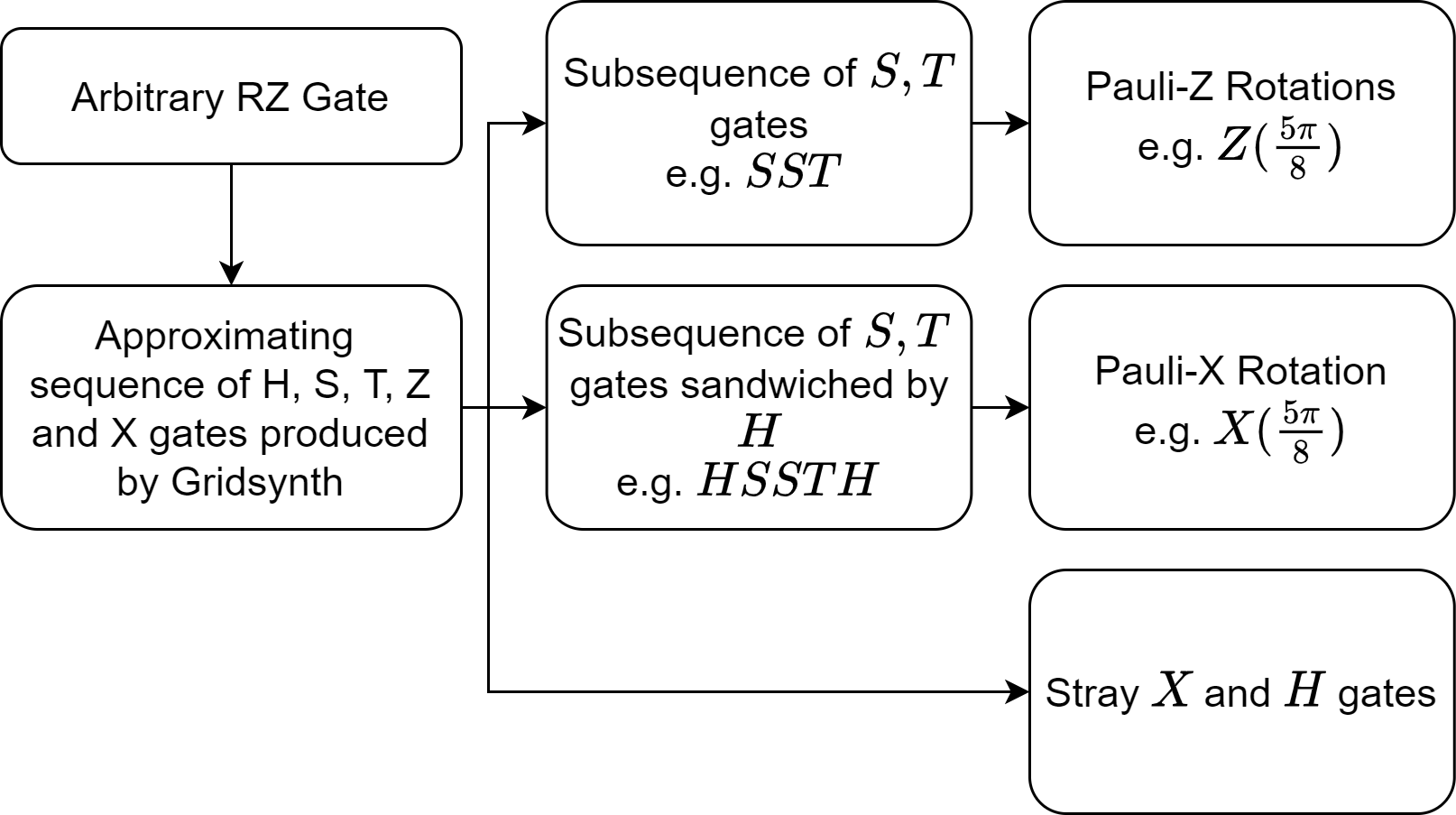}
    \caption{Pauli rotation compression of Cliffod+T approximations of small angle rotations E.g. $R_Z(\frac{\pi}{2^{128}})$: gate sequences obtained from Gridsynth are interpreted in the Pauli Frame by breaking it into subsequences. For instance, the sequence $HSHTSHX$ would be split as $HSH$, $TS$, $H$, $X$ and would become the sequence $X_{\frac{\pi}{4}} Z_{\frac{3\pi}{8}} HX$}
    \label{fig:gridsynthToRot}
\end{figure}

We utilize two methods to convert the Gridsynth appproximation to LLI. The first is to directly apply the gates with the methods of Table~\ref{table:SurfaceCodeOp}: H, X and Z transversally, S with a twist and T as $Z(\frac{\pi}{8})$ rotation as shown in Fig.~\ref{fig:GatesToLLI}. The second approach, we refer to as \emph{Pauli rotation compression}, is shown in Fig.~\ref{fig:gridsynthToRot}, and consists of interpreting the gate sequences returned by the Gridsynth approximations as a sequence of Pauli rotations of varying angles.

The direct application of gates is simpler and results in the same Clifford corrective terms for every rotation. With Pauli rotation compression the Clifford corrective terms change for every angle, thus more complex classical control would be required by a downstream stage. In the Appendix we present an algorithm for Clifford gate optimization.

\subsection{Slices and Routing}
\label{sec:routing}

\begin{figure}[h]
    \centering
    \includegraphics[width=0.55\columnwidth]{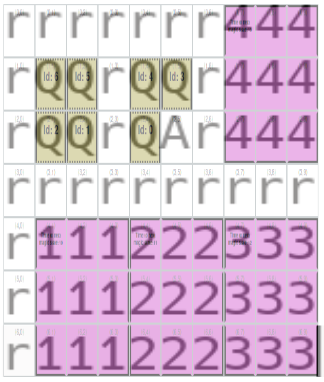}
    \caption{ASCII specification for patch layouts. \code{Q} indicates a patch holding a logical qubit, \code{r} marks cells that are reserved for routing (the cyan ``snakes" of Fig.~\ref{fig:LargeScaleComputation}). Numbers \code{0} to \code{9} are used to identify distillation regions. The boundaries of the distillation regions are computed by connected components search for same numbers, so it is possible to have more than 10 distillation regions. Magic states produced by these regions are queued in the \code{r} cells neighbouring a distillation block. Finally, \code{A} marks cells reserved to allocate new ancillae patches in the states, such as the $\ket{+}$ states that mediate CNOTs or the places for the $Y$ eigenstates used by $\frac{n}{4}$ Pauli rotations. The layout file format is described in the Appendix.}
    \label{fig:layout}
\end{figure}

To overcome the logistical challenges of structuring a computation on surface code device, we arrange the computation in space and time. Space structure is given by partitioning the physical lattice into square \textit{cells}. A cell may hold or may not hold a patch, be part of a distillation region, or may be used for routing, but patches, distillation regions and routing areas are always placed in accordance to cell boundaries ( Figs.~\ref{fig:LargeScaleComputation}, \ref{fig:singlepatches} and \ref{fig:layout}).

Time structure is given by thinking of the computation in terms of \textit{slices}. Surface code computations can be viewed as 3D structures in space-time~\cite{paler2016synthesis, paler2019opensurgery, gidney2019flexible}, and a slice is a plane through the structure at a fixed time value (Fig.~\ref{fig:LargeScaleComputation}). In a nutshell, a slice is a temporally discretized partition of the computation (\textit{clock timesteps} in Litinski~\cite{litinski2019game} or \emph{moments} in Google Cirq~\cite{cirq_developers_2021_5182845} terminology, for example). Each slice represents a snapshot of the the LLIs that are happening simultaneously on the lattice -- slice duration is given by the duration of the slowest LLI.

\emph{Routing} is the problem of deciding how the cells of a slice are allocated to patches or reserved for other purposes. Finding optimal layouts has a great impact on the depth of the computation. Different layouts can for example be used to trade off space for time~\cite{lao2018mapping}. Layouts may need to change depending on the task during an algorithm. For example, the oracle in Grover's search algorithm  may be very different from the implementation of the diffusion operator~\cite{jaques2020implementing}. We defined our own \textit{configurable layout specification} (Fig.~\ref{fig:layout}). The compiler reads a text file containing the layout specification and produces slices with patches arranged accordingly.

\subsubsection{On-the-fly, Functionally Verified Slicer}
\label{sec:slowSlicer}

Our first slicer supports the real-time, on-the-fly functional verification for correctness. This slicer can be used as a preliminary verification of smaller scale lattice surgery circuits. The slicer and the simulation operate on an array of patches of variable length and assumes that all magic states have been prepared ahead of time. The verified slicer is very powerful when it comes to understanding the details of small computation and we used it in the development of the compiler.

The simulator, called the \emph{Lazily Tensored State-vector Simulation} (LTSvS), has the major feature of being able to simulate patch states at the LLI instruction level, such as simulating multi body measurements and Pauli operator gates. LTSvS tensors at the matrix level only when strictly required, otherwise just tracks the fact that the global state is given by a tensor product of sub vectors. LTSvS offers a great performance advantage over n\"aive state vector simulation: our simulator doesn't expand the full state vector of all logical qubits on the lattice. In particular, qubits that are known to not be entangled, because they were just initialized or measured, are automatically tracked in separate sub-state vectors. Qubits may be entangled within a sub-state vector. An example of unentangled qubits is the array of magic states waiting to be used or ancillae patches.

Methodologically, the LTSvS simulator is very similar to the matrix-product-states (MPS) simulation techniques~\cite{vidal2003efficient}, which are efficient on circuits with low counts of entangling gates. Compared to the MPS simulators, e.g. from Qiskit~\cite{Qiskit}, ours is fine tuned for computations with many ancillae and measurements, can handle classical control and can be executed in parallel with the compilation process.

\subsubsection{High Performance Slicer}
\label{sec:cppSlicer}

The main goal of our compiler is to handle very large scale circuits with thousands of logical qubits and millions of LLIs. At this scale every CPU clock cycle and every byte are precious. Our high performance compiler is written in C++ because it comes with zero cost abstraction~\cite{stroustrup2004abstraction}.

The first step of the slicer is to read a layout file (Fig.~\ref{fig:layout}) in order to create an abstract layout representation that describes the device layout. The layout is used to initialize a slice template which will be reused for the routing. The template will be recomputed once the layout dictates this. Our implementation of slice processing keeps memory usage to a minimum because $O(1)$ slices are ever kept in memory by the slicer itself. Moreover, this representation is stored in a high performance data structure based on bitstreams and hash-maps. The representation will be used for computing routes using a variant of Dijkstra's algorithm. 

The slicer streams LLI from text or standard input, updating the slice with each instruction, evolving the slices over time. Since the slicer can also stream read from standard input and write to standard output, its possible to implement external programs (e.g. Python scripts or other command line tools) that visit slices by reading from standard input. Given the capability of evolving the lattice state, the slice processing functionality is implemented by defining a C++ functor to visit all slices.

The streamed evolution of slices includes managing distillation, queuing magic states~\cite{paler2016synthesis, paler2017synthesissystematic}, initializing ancillae and LLI operations. A user may collect statistics on slices, such as magic state queue and routing space usage in seconds, without having to store in terabytes of memory that slow down the processing. At the same time this functor approach has the advantage of hiding the implementation details from the client so that they can focus on the processing functionality.

To place routing regions, we used our own implementation of Dijkstra's algorithm, which is implemented \emph{in place}, such that our tool can search the lattice without constructing a graph of it. Our implementation has close to zero cost overheads with respect to memory and CPU instructions needed to translate back and forth between the lattice layout and the graph needed for performing Dijkstra's algorithm. To further speed up routing, we employ a \emph{cached routing} technique where previously computed routes are saved and reused later.

\section{Results}
\label{sec:res}

We present results for compiling very large quantum circuits, and focus on scalability and resource estimation.

\begin{figure}[!t]
    \centering
    \includegraphics[width=\columnwidth]{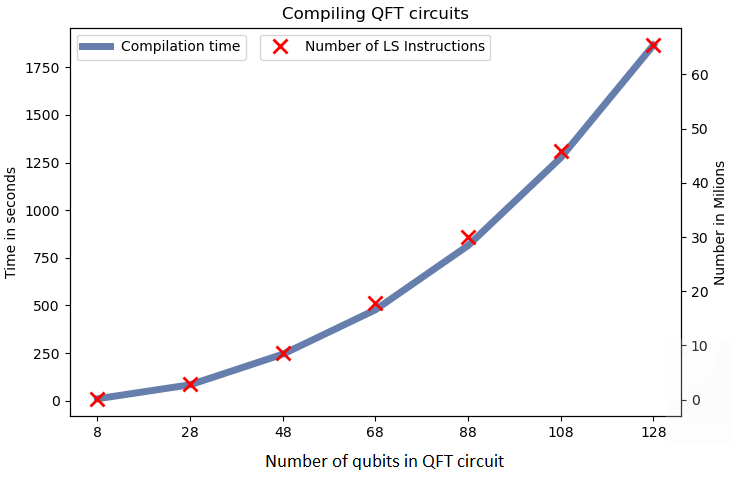}
    \caption{Time taken to compile a QFT with the C++ slicer on a laptop (Intel i5350U, 8GB RAM) and number of LLI instructions for different QFT sizes. The Clifford+T implementation of the QFT requires thousands of gates for each controlled rotation, to retain the rotation accuracy we set ($10^{-41}$).}
    \label{fig:QFTBenchmark}
\end{figure}

\subsection{128-Qubit QFT}

To validate the performance of our compiler and high performance slicer we took a circuit that has wide spread use and presents technical fault-tolerant execution challenges. The Quantum Fourier Transform (QFT) is a crucial component providing quantum speedup to algorithms such as Shor's algorithm and quantum phase estimation~\cite{nielsen2001quantum}. The fault-tolerant implementation of the QFT is challenging because of the presence of small angle controlled rotations. For the QFT to retain the desired level of precision, these have to be approximated by a long sequence of Clifford+T, which results in a very long computation. We set Gridsynth's precision to $10^{-41}$ for the Clifford+T approximations for small angle rotations, which results in thousands of gates. Such number was chosen as it is 3 orders of magnitude less than the smallest angle rotation in our circuit $\frac{\pi}{2^{128}} \approx 10^{-38}$, after expanding out the controlled rotations (Sec.~\ref{sec:TransformingTheGateSet}).

The number of controlled rotations increases quadratically with the number of qubits the QFT is applied to. Thus, at 128 qubits and after small angle rotation approximation, the QFT circuit has more than 80 Million LLI without gate to Pauli compression. The number of LLI includes Clifford corrective terms that are meant to be applied depending on measurement outcomes. Thanks to concurrent magic state distillation, there are no idle slices waiting for magic states to be produced. We used the high performance slicer to compile the 128-qubit QFT: for example, laying out the slices for the roughly 80 million LLI of the 128-qubit QFT takes less than 15 minutes on an ordinary laptop. The generation of LLI of a QFT on 128 qubits takes negligible time (under 10s on a laptop). Fig.~\ref{fig:QFTBenchmark} illustrates the performance of the C++ slicer for the QFT128 circuit.

\begin{figure}[!t]
    \centering
    \includegraphics[width=\columnwidth]{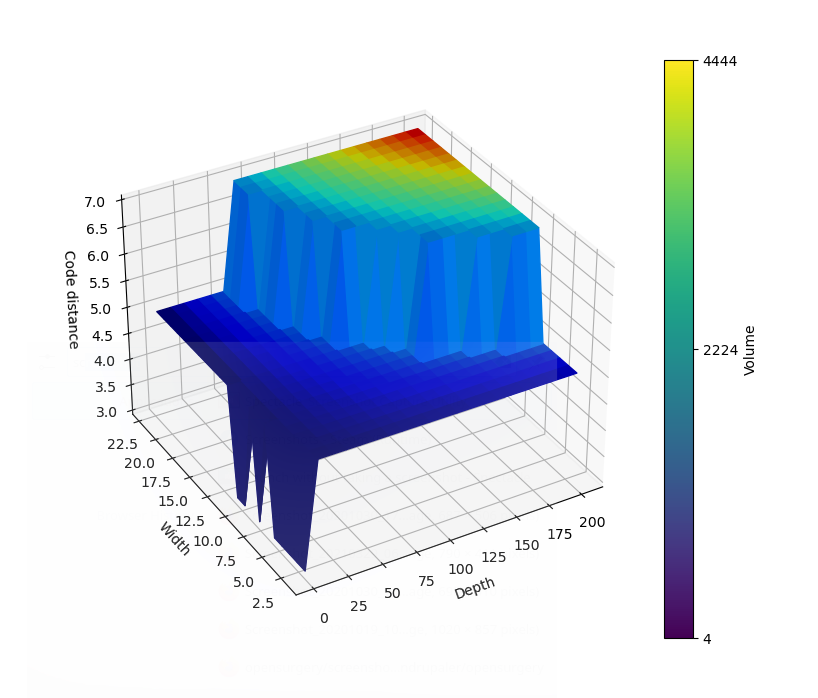}
    \caption{A snapshot of the resource requirement landscape for random H, T and CNOT circuits. The horizontal axes show circuit width (number of qubits) and depth (number of gates). The vertical axis shows the code distance required to execute the desired circuit with a success rate of 99\%. The colour scale represents the space-time volume of the computation, which relates closely with code distance.}
    \label{fig:Resources}
\end{figure}

\subsection{Resource Estimation}
\label{sec:qentian}

A challenging problem in the community of fault-tolerant QC is determining the amount of physical resource that is necessary to carry out a logical computation with a certain degree of precision. Such resource is often quantified by  physical qubits over time -- often called a \emph{space time volume}. The depth of the circuit and the required magic state fidelity affect the code distance, which in turn affect the number of physical qubits required. Moreover, the degree of parallelization achieved at the routing stage will affect computation depth.

Our compiler includes a prototypical resource estimator for surface code computations. We use the Qentiana~\cite{herr2017lattice} software to estimate such values, and computed some code distances for randomized circuits of H, T and CNOT gates (Fig.~\ref{fig:Resources}).

\section{Conclusion}

We introduced and described a compiler for lattice surgery quantum circuits and showcased some of the results achieved with it. We motivated the design choices behind our two stage pipeline. The first stage included how input circuits are parsed, pre-processed, reduced to Clifford+T and viewed as Pauli rotations. The second stage focuses on laying out circuits on physical devices, which presents substantial performance challenges.

We demonstrated the compiler's performance by compiling a 128-qubit QFT. We believe this is a notable achievement: despite its widespread appearance in algorithms, to the best of our knowledge, no surface code compiler is able to handle such large scale circuits. We also showcased the compiler's ability to estimate resource requirements, in particular patch code distance, which is promising in the perspective of quantum benchmarking.

Our project is laying the foundation for a full-stack quantum circuit compilation framework. Future work will leverage hybrid classical and quantum instruction sets such as LLVM/QIR to program high performance classical control while integrating the Quantum Processing Unit (QPU) instructions.

\section*{Acknowledgements}
We thank Tyler LeBond for contributing important code supporting the improved layout functionality, and Niki Loppi of the NVIDIA AI Technology Center Finland for his help with the implementation. We thank  Varun Seshadri for his feedback and helpful discussions.

George Watkins and Alexandru Paler were with funding from the Defense Advanced Research Projects Agency [under the Quantum Benchmarking (QB) program under award no. HR00112230007 and HR001121S0026 contracts]. The views, opinions and/or findings expressed are those of the authors and should not be interpreted as representing the official views or policies of the Department of Defense or the U.S. Government. Hoang Minh Nguyen, Keelan Watkins and George Watkins have been supported by the Unitary.fund. Hoi-Kwan Lau acknowledges support from the Canada Research Chair. Alexandru Paler acknowledges a Google Gift 2023, and funding received from the Finnish-American Research and Innovation Accelerator, one of eight global pilots funded by the Finnish Ministry of Education and Culture.

\bibliographystyle{quantum}
\bibliography{__cites}

\begin{thebibliography}{10}

\bibitem{suchara2013comparing}
Martin Suchara, John Kubiatowicz, Arvin Faruque, Frederic~T. Chong, Ching-Yi
  Lai, and Gerardo Paz.
\newblock ``Qure: The quantum resource estimator toolbox''.
\newblock In 2013 IEEE 31st International Conference on Computer Design (ICCD).
\newblock \href{https://dx.doi.org/10.1109/ICCD.2013.6657074}{Pages 419--426}.
\newblock ~(2013).

\bibitem{preskill2018quantumcomputingin}
John Preskill.
\newblock ``Quantum {C}omputing in the {NISQ} era and beyond''.
\newblock \href{https://dx.doi.org/10.22331/q-2018-08-06-79}{{Quantum} {\bf 2},
  79}~(2018).

\bibitem{parker2000efficient}
S.~Parker and M.~B. Plenio.
\newblock ``Efficient factorization with a single pure qubit and
  $\mathrm{log}\mathit{N}$ mixed qubits''.
\newblock \href{https://dx.doi.org/10.1103/PhysRevLett.85.3049}{Phys. Rev.
  Lett. {\bf 85}, 3049--3052}~(2000).

\bibitem{gidney2021factor}
Craig Gidney and Martin Eker{\aa{}}.
\newblock ``How to factor 2048 bit {RSA} integers in 8 hours using 20 million
  noisy qubits''.
\newblock \href{https://dx.doi.org/10.22331/q-2021-04-15-433}{{Quantum} {\bf
  5}, 433}~(2021).

\bibitem{arute2019quantum}
Frank Arute, Kunal Arya, Ryan Babbush, Dave Bacon, Joseph~C Bardin, Rami
  Barends, Rupak Biswas, Sergio Boixo, Fernando~GSL Brandao, David~A Buell,
  et~al.
\newblock ``Quantum supremacy using a programmable superconducting processor''.
\newblock \href{https://dx.doi.org/10.1038/s41586-019-1666-5}{Nature {\bf 574},
  505--510}~(2019).

\bibitem{fowler2012surface}
Austin~G Fowler, Matteo Mariantoni, John~M Martinis, and Andrew~N Cleland.
\newblock ``Surface codes: Towards practical large-scale quantum computation''.
\newblock \href{https://dx.doi.org/10.1103/PhysRevA.86.032324}{Physical Review
  A {\bf 86}, 032324}~(2012).

\bibitem{wang2011surface}
David~S Wang, Austin~G Fowler, and Lloyd~CL Hollenberg.
\newblock ``Surface code quantum computing with error rates over 1\%''.
\newblock \href{https://dx.doi.org/10.1103/PhysRevA.83.020302}{Physical Review
  A {\bf 83}, 020302}~(2011).

\bibitem{tomita2014low}
Yu~Tomita and Krysta~M Svore.
\newblock ``Low-distance surface codes under realistic quantum noise''.
\newblock \href{https://dx.doi.org/10.1103/PhysRevA.90.062320}{Physical Review
  A {\bf 90}, 062320}~(2014).

\bibitem{andersen2020correction}
Sebastian Krinner, Nathan Lacroix, Ants Remm, Agustin Di~Paolo, Elie Genois,
  Catherine Leroux, Christoph Hellings, Stefania Lazar, Francois Swiadek,
  Johannes Herrmann, Graham~J. Norris, Christian~Kraglund Andersen, Markus
  M\"{u}ller, Alexandre Blais, Christopher Eichler, and Andreas Wallraff.
\newblock ``Realizing repeated quantum error correction in a distance-three
  surface code''~(2021).

\bibitem{andersen2020detection}
Christian~Kraglund Andersen, Ants Remm, Stefania Lazar, Sebastian Krinner,
  Nathan Lacroix, Graham~J. Norris, Mihai Gabureac, Christopher Eichler, and
  Andreas Wallraff.
\newblock ``Repeated quantum error detection in a surface code''.
\newblock \href{https://dx.doi.org/10.1038/s41567-020-0920-y}{Nature Physics
  {\bf 16}, 875--880}~(2020).

\bibitem{acharya2022suppressing}
Rajeev Acharya, Igor Aleiner, Richard Allen, Trond~I Andersen, Markus Ansmann,
  Frank Arute, Kunal Arya, Abraham Asfaw, Juan Atalaya, Ryan Babbush, et~al.
\newblock ``Suppressing quantum errors by scaling a surface code logical
  qubit''.
\newblock \href{https://dx.doi.org/10.1038/s41586-022-05434-1}{Nature {\bf
  614}, 676--681}~(2023).

\bibitem{bourassa2021blueprint}
J.~Eli Bourassa, Rafael~N. Alexander, Michael Vasmer, Ashlesha Patil, Ilan
  Tzitrin, Takaya Matsuura, Daiqin Su, Ben~Q. Baragiola, Saikat Guha, Guillaume
  Dauphinais, Krishna~K. Sabapathy, Nicolas~C. Menicucci, and Ish Dhand.
\newblock ``Blueprint for a {S}calable {P}hotonic {F}ault-{T}olerant {Q}uantum
  {C}omputer''.
\newblock \href{https://dx.doi.org/10.22331/q-2021-02-04-392}{{Quantum} {\bf
  5}, 392}~(2021).

\bibitem{hanks2020effective}
Michael Hanks, Marta~P. Estarellas, William~J. Munro, and Kae Nemoto.
\newblock ``Effective compression of quantum braided circuits aided by
  zx-calculus''.
\newblock \href{https://dx.doi.org/10.1103/PhysRevX.10.041030}{Phys. Rev. X
  {\bf 10}, 041030}~(2020).

\bibitem{gidney2019flexible}
Craig Gidney and Austin~G. Fowler.
\newblock ``Flexible layout of surface code computations using autoccz
  states''~(2019).
\newblock  \href{http://arxiv.org/abs/1905.08916}{arXiv:1905.08916}.

\bibitem{herr2017optimization}
Daniel Herr, Franco Nori, and Simon~J Devitt.
\newblock ``Optimization of lattice surgery is np-hard''.
\newblock \href{https://dx.doi.org/10.1038/s41534-017-0035-1}{npj Quantum
  Information {\bf 3}, 35}~(2017).

\bibitem{wasa2023hardness}
Kunihiro Wasa, Shin Nishio, Koki Suetsugu, Michael Hanks, Ashley Stephens,
  Yu~Yokoi, and Kae Nemoto.
\newblock ``Hardness of braided quantum circuit optimization in the surface
  code''.
\newblock \href{https://dx.doi.org/10.1109/TQE.2023.3251358}{IEEE Transactions
  on Quantum Engineering {\bf 4}, 1--7}~(2023).

\bibitem{paler2020aggregated}
Alexandru Paler.
\newblock ``Aggregated control of quantum computations: When stacked
  architectures are too good to be practical soon''.
\newblock \href{https://dx.doi.org/10.1109/MC.2020.2997277}{Computer {\bf 53},
  74--78}~(2020).

\bibitem{nielsen2001quantum}
Michael~A. Nielsen and Isaac~L. Chuang.
\newblock ``Quantum computation and quantum information''.
\newblock Cambridge University Press. ~(2000).

\bibitem{mermin2007quantum}
N~David Mermin.
\newblock ``Quantum computer science: an introduction''.
\newblock Cambridge University Press. ~(2007).

\bibitem{varsamopoulos2020decoding}
Savvas Varsamopoulos, Koen Bertels, and Carmen~G. Almudever.
\newblock ``Decoding surface code with a distributed neural network\textendash
  based decoder''.
\newblock \href{https://dx.doi.org/10.1007/s42484-020-00015-9}{Quantum Machine
  Intelligence {\bf 2}, 1--12}~(2020).

\bibitem{hu2020quasilinear}
Mark~Shui Hu and David Elkouss.
\newblock ``Quasilinear time decoding algorithm for topological codes with high
  error threshold''.
\newblock \href{https://dx.doi.org/10.13140/RG.2.2.13495.96162}{Master's
  thesis, TU Delft}~(2020).

\bibitem{terhal2015quantum}
Barbara~M. Terhal.
\newblock ``Quantum error correction for quantum memories''.
\newblock \href{https://dx.doi.org/10.1103/RevModPhys.87.307}{Rev. Mod. Phys.
  {\bf 87}, 307--346}~(2015).

\bibitem{roffe2019quantum}
Joschka Roffe.
\newblock ``Quantum error correction: an introductory guide''.
\newblock \href{https://dx.doi.org/10.1080/00107514.2019.1667078}{Contemporary
  Physics {\bf 60}, 226--245}~(2019).

\bibitem{kitaev2003fault}
A.Yu. Kitaev.
\newblock ``Fault-tolerant quantum computation by anyons''.
\newblock
  \href{https://dx.doi.org/https://doi.org/10.1016/S0003-4916(02)00018-0}{Annals
  of Physics {\bf 303}, 2--30}~(2003).

\bibitem{googleResearchJourney}
Google Research.
\newblock ``Google quantum ai journey''~(2022).

\bibitem{IBMResearchRoadmap}
IBM Research.
\newblock ``Ibm quantum roadmap''~(2022).

\bibitem{bravyi1998quantum}
S.~B. Bravyi and A.~Yu. Kitaev.
\newblock ``Quantum codes on a lattice with boundary''~(1998).
\newblock
  \href{http://arxiv.org/abs/quant-ph/9811052}{arXiv:quant-ph/9811052}.

\bibitem{dennis2001topological}
Eric Dennis, Alexei Kitaev, Andrew Landahl, and John Preskill.
\newblock ``{Topological quantum memory}''.
\newblock \href{https://dx.doi.org/10.1063/1.1499754}{Journal of Mathematical
  Physics {\bf 43}, 4452--4505}~(2002).

\bibitem{horsman2012surface}
Dominic Horsman, Austin~G Fowler, Simon Devitt, and Rodney~Van Meter.
\newblock ``Surface code quantum computing by lattice surgery''.
\newblock \href{https://dx.doi.org/10.1088/1367-2630/14/12/123011}{New Journal
  of Physics {\bf 14}, 123011}~(2012).

\bibitem{poulsen2017fault}
Hendrik Poulsen~Nautrup, Nicolai Friis, and Hans~J Briegel.
\newblock ``Fault-tolerant interface between quantum memories and quantum
  processors''.
\newblock \href{https://dx.doi.org/10.1038/s41467-017-01418-2}{Nature
  communications {\bf 8}, 1--8}~(2017).

\bibitem{litinski2019game}
Daniel Litinski.
\newblock ``A game of surface codes: Large-scale quantum computing with lattice
  surgery''.
\newblock \href{https://dx.doi.org/10.22331/q-2019-03-05-128}{Quantum {\bf 3},
  128}~(2019).

\bibitem{herr2019time}
Daniel Herr, Alexandru Paler, Simon~J. Devitt, and Franco Nori.
\newblock ``Time versus hardware: Reducing qubit counts with a (surface code)
  data bus''~(2019).
\newblock  \href{http://arxiv.org/abs/1902.08117}{arXiv:1902.08117}.

\bibitem{bravyi2005universal}
Sergey Bravyi and Alexei Kitaev.
\newblock ``Universal quantum computation with ideal clifford gates and noisy
  ancillas''.
\newblock \href{https://dx.doi.org/10.1103/PhysRevA.71.022316}{Phys. Rev. A
  {\bf 71}, 022316}~(2005).

\bibitem{bravyi2012magic}
Sergey Bravyi and Jeongwan Haah.
\newblock ``Magic-state distillation with low overhead''.
\newblock \href{https://dx.doi.org/10.1103/PhysRevA.86.052329}{Phys. Rev. A
  {\bf 86}, 052329}~(2012).

\bibitem{litinski2019magic}
Daniel Litinski.
\newblock ``Magic state distillation: Not as costly as you think''.
\newblock \href{https://dx.doi.org/10.22331/q-2019-12-02-205}{Quantum {\bf 3},
  205}~(2019).

\bibitem{paler2019surfbraid}
Alexandru Paler.
\newblock ``Surfbraid: A concept tool for preparing and resource estimating
  quantum circuits protected by the surface code''~(2019).
\newblock  \href{http://arxiv.org/abs/1902.02417}{arXiv:1902.02417}.

\bibitem{paler2019opensurgery}
Alexandru Paler and Austin~G. Fowler.
\newblock ``Opensurgery for topological assemblies''.
\newblock In 2020 IEEE Globecom Workshops (GC Wkshps.
\newblock \href{https://dx.doi.org/10.1109/GCWkshps50303.2020.9367489}{Pages
  1--4}.
\newblock ~(2020).

\bibitem{lao2018mapping}
Lingling Lao, Bas van Wee, Imran Ashraf, J~van Someren, Nader Khammassi, Koen
  Bertels, and Carmen~G Almudever.
\newblock ``Mapping of lattice surgery-based quantum circuits on surface code
  architectures''.
\newblock \href{https://dx.doi.org/10.1088/2058-9565/aadd1a}{Quantum Science
  and Technology {\bf 4}, 015005}~(2018).

\bibitem{hua2021autobraid}
Fei Hua, Yanhao Chen, Yuwei Jin, Chi Zhang, Ari Hayes, Youtao Zhang, and
  Eddy~Z. Zhang.
\newblock ``Autobraid: A framework for enabling efficient surface code
  communication in quantum computing''.
\newblock In MICRO-54: 54th Annual IEEE/ACM International Symposium on
  Microarchitecture.
\newblock \href{https://dx.doi.org/10.1145/3466752.3480072}{Page 925–936}.
\newblock MICRO '21New York, NY, USA~(2021). Association for Computing
  Machinery.

\bibitem{beverland2021surface}
Michael Beverland, Vadym Kliuchnikov, and Eddie Schoute.
\newblock ``Surface code compilation via edge-disjoint paths''.
\newblock \href{https://dx.doi.org/10.1103/PRXQuantum.3.020342}{PRX Quantum
  {\bf 3}, 020342}~(2022).

\bibitem{xu2019variational}
Xiaosi Xu, Simon~C. Benjamin, and Xiao Yuan.
\newblock ``Variational circuit compiler for quantum error correction''.
\newblock \href{https://dx.doi.org/10.1103/PhysRevApplied.15.034068}{Phys. Rev.
  Appl. {\bf 15}, 034068}~(2021).

\bibitem{litinski2018lattice}
Daniel Litinski and Felix von Oppen.
\newblock ``Lattice surgery with a twist: Simplifying clifford gates of surface
  codes''.
\newblock \href{https://dx.doi.org/10.22331/q-2018-05-04-62}{Quantum {\bf 2},
  62}~(2018).

\bibitem{cross2017open}
Andrew~W. Cross, Lev~S. Bishop, John~A. Smolin, and Jay~M. Gambetta.
\newblock ``Open quantum assembly language''~(2017).
\newblock  \href{http://arxiv.org/abs/1707.03429}{arXiv:1707.03429}.

\bibitem{Qiskit}
H~Abraham et~al.
\newblock ``Qiskit: An open-source framework for quantum computing''~(2021).

\bibitem{kissinger2020Pyzx}
Aleks Kissinger and John van~de Wetering.
\newblock ``{PyZX: Large Scale Automated Diagrammatic Reasoning}''.
\newblock In Bob Coecke and Matthew Leifer, editors, {\rm Proceedings 16th
  International Conference on} Quantum Physics and Logic, {\rm Chapman
  University, Orange, CA, USA., 10-14 June 2019}.
\newblock \href{https://dx.doi.org/10.4204/EPTCS.318.14}{Volume 318 of
  Electronic Proceedings in Theoretical Computer Science, pages 229--241}.
\newblock Open Publishing Association~(2020).

\bibitem{jochym2018disjointness}
Tomas Jochym-O'Connor, Aleksander Kubica, and Theodore~J. Yoder.
\newblock ``Disjointness of stabilizer codes and limitations on fault-tolerant
  logical gates''.
\newblock \href{https://dx.doi.org/10.1103/PhysRevX.8.021047}{Phys. Rev. X {\bf
  8}, 021047}~(2018).

\bibitem{ross2014optimal}
Neil~J. Ross and Peter Selinger.
\newblock ``Optimal ancilla-free clifford+t approximation of z-rotations''.
\newblock Quantum Info. Comput. {\bf 16}, 901–953~(2016).

\bibitem{paler2016synthesis}
Alexandru Paler, Simon~J Devitt, and Austin~G Fowler.
\newblock ``Synthesis of arbitrary quantum circuits to topological assembly''.
\newblock \href{https://dx.doi.org/10.1038/s41598-017-10657-8}{Scientific
  reports {\bf 6}, 1--16}~(2016).

\bibitem{cirq_developers_2021_5182845}
Cirq Developers.
\newblock ``Cirq''.
\newblock \href{https://dx.doi.org/10.5281/zenodo.5182845}{Zenodo}~(2021).

\bibitem{jaques2020implementing}
Samuel Jaques, Michael Naehrig, Martin Roetteler, and Fernando Virdia.
\newblock ``Implementing grover oracles for quantum key search on aes and
  lowmc''.
\newblock In Advances in Cryptology – EUROCRYPT 2020: 39th Annual
  International Conference on the Theory and Applications of Cryptographic
  Techniques, Zagreb, Croatia, May 10–14, 2020, Proceedings, Part II.
\newblock \href{https://dx.doi.org/10.1007/978-3-030-45724-2_10}{Page
  280–310}.
\newblock Berlin, Heidelberg~(2020). Springer-Verlag.

\bibitem{vidal2003efficient}
Guifr\'e Vidal.
\newblock ``Efficient classical simulation of slightly entangled quantum
  computations''.
\newblock \href{https://dx.doi.org/10.1103/PhysRevLett.91.147902}{Phys. Rev.
  Lett. {\bf 91}, 147902}~(2003).

\bibitem{stroustrup2004abstraction}
Bjarne Stroustrup.
\newblock ``Keynote speech: Abstraction and the c++ machine model''.
\newblock In Proceedings of the First International Conference on Embedded
  Software and Systems.
\newblock \href{https://dx.doi.org/10.1007/11535409_1}{Page 1–13}.
\newblock ICESS'04Berlin, Heidelberg~(2004). Springer-Verlag.

\bibitem{paler2017synthesissystematic}
Alexandru Paler, Austin~G Fowler, and Robert Wille.
\newblock ``Synthesis of arbitrary quantum circuits to topological assembly:
  Systematic, online and compact''.
\newblock \href{https://dx.doi.org/10.1038/s41598-017-10657-8}{Scientific
  reports {\bf 7}, 1--16}~(2017).

\bibitem{herr2017lattice}
Daniel Herr, Franco Nori, and Simon~J Devitt.
\newblock ``Lattice surgery translation for quantum computation''.
\newblock \href{https://dx.doi.org/10.1088/1367-2630/aa5709}{New Journal of
  Physics {\bf 19}, 013034}~(2017).

\bibitem{jq}
JQ-Authors.
\newblock ``Command-line {JSON} processor''.
\newblock \url{https://github.com/stedolan/jq}~(2022).

\bibitem{kim2018gateidentity}
Taewan Kim and Byung-Soo Choi.
\newblock ``Efficient decomposition methods for controlled-r n using a single
  ancillary qubit''.
\newblock \href{https://dx.doi.org/10.1038/s41598-018-23764-x}{Scientific
  Reports{\bf 8}}~(2018).

\bibitem{aaronson2004improved}
Scott Aaronson and Daniel Gottesman.
\newblock ``Improved simulation of stabilizer circuits''.
\newblock \href{https://dx.doi.org/10.1103/PhysRevA.70.052328}{Phys. Rev. A
  {\bf 70}, 052328}~(2004).

\end{thebibliography}

\section*{Appendix}

\subsection{OpenQASMMin}
\label{sec:qasmmin}

The natively supported subset of OpenQASM 2.0 instructions is presented in Table~\ref{tab:qasmmin}. Our compiler can parse a small subset of OpenQASM 2.0 instead of LLI. We call this type of assembly OpenQASMmin. In general, OpenQASMmin should be valid OpenQASM, with the restrictions below:
\begin{itemize}
    \setlength{\itemsep}{0pt}
    \setlength{\parskip}{0pt}
    \item Program must begin with \texttt{OPENQASM 2.0;}
    \item Only one register is allowed (whether the names match will not be checked)
    \item At most one gate per line
    \item Single qubit gates must be in the form \texttt{g q[n]}; where \texttt{g} is one of \texttt{h,x,z,s,t} and \texttt{n} is a non-negative integer
    \item \texttt{rz(expr)} and \texttt{crz(expr)} where \texttt{expr} has form \texttt{pi/m} or \texttt{n*pi/m} for \texttt{n}, m integers. No whitespace.
    \item CNOTs must be in the form \texttt{cx q[n],q[m];} where \texttt{n} and \texttt{m} are non-negative, and target is first;
    \item No classical control is supported;
    \item No measurement operators are supported;
    \item Only inline comments, e.g. \texttt{cx q[0],q[7]; //cnot on 0 and 7}.
\end{itemize}

\begin{table}[!h]
  \centering
    \caption{OpenQASMMin: Supported instructions}
    \label{tab:qasmmin}
    
  \begin{tabular}{|p{.3\columnwidth}| p{.6\columnwidth}|}
    \hline
        \multicolumn{2}{|p{.9\columnwidth}|}{Classical registers, barriers and include directives are ignored.}\\
    \hline
        \multicolumn{2}{|p{.9\columnwidth}|}{Unsupported instructions and gates raise an exception.} \\
    \hline
        \code{h}, \code{x}, \code{z}, \code{s}, \code{sdg}, \code{t}, \code{tdg} & Single qubit gates with the usual meaning. The \code{-dg} suffix stands for dagger.\\
    \hline
        \code{cnot} or \code{cx}, \code{cz} & Controlled X and Z gates. \\ 
    \hline
        \code{rx(theta)}, \code{rz(theta)} & $R_Z$ and $R_X$ gates. Argument \code{theta} has to have form \code{[N*pi/D} for integers \code{N} and \code{D}, with the \code{D} being a non negative power of two. Both \code{N} and \code{D} can be of arbitrary size. \\
    \hline
        \code{crx(theta)}, \code{crz(theta)} & Controlled $R_Z$ and $R_X$ gates. The argument has the same form as for the $R_Z$ and $R_X$ gates. \\
    \hline
        \code{qreg} & Only one quantum register is supported. \\
    \hline
  \end{tabular}
\end{table}

\subsection{Compiler Pipeline and Operating Systems}

It is possible to accept a wider range of circuits with additional processing by using Qiskit and PyZX. For example, we have successfully processed Grover circuits with multi qubit gates and Toffoli based adders by decomposing additional gates such as \code{mcx}, \code{ccx} and \code{rccx} which are not in the natively supported set. This kind of gate decomposition is done on a case by case basis.

\begin{figure}[!h]
    \centering
    \includegraphics[width=\columnwidth]{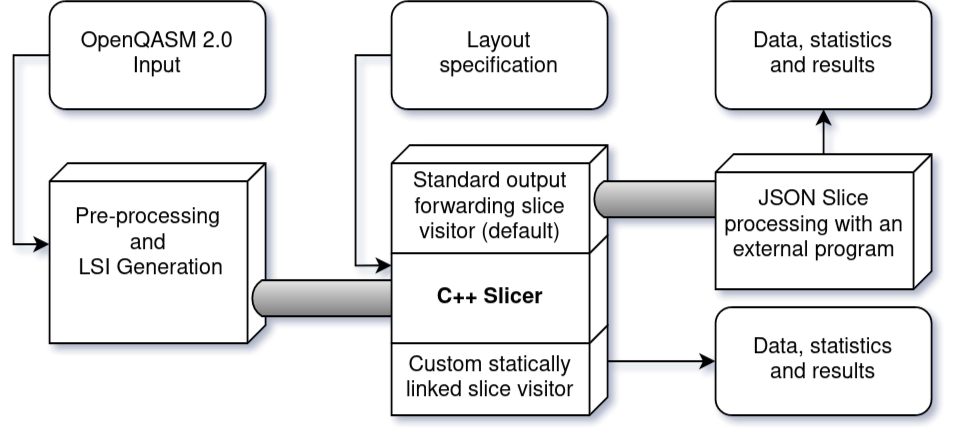} \\
    E.g.: \code{\$./qft\_to\_stdout.py 128 | lsqecc\_slicer -l 12by12.txt | jq .[][][] | grep null | wc -l
}
    \caption{ \code{qft\_to\_stdout.py} is our tool to stream LLI for a high precision QFT, the \code{128} argument indicates the number of qubits. \code{lsqecc\_slicer -f json} is our C++ slicer, which lays out the LLI instructions and outputs a stream of JSON slices to stout, the \code{-l 12by12.txt} flag tells it which layout to use. Finally \code{jq}~\cite{jq} is a streaming JSON processor that can extract information from the sequence of slices, here combined with some POSIX utilities to count the total number of routing cells.}
    \label{fig:pipedSliceing}
\end{figure}

Figure~\ref{fig:pipedSliceing} is a diagram of how the C++ slicer can take advantage of the operating systems ability to broker messages by using POSIX pipes and an example shell command to run such a pipeline.

\subsection{Circuit Simulation and Gate Decomposition}

\begin{figure}[!h]
    \centering
    \includegraphics[width=0.95\columnwidth]{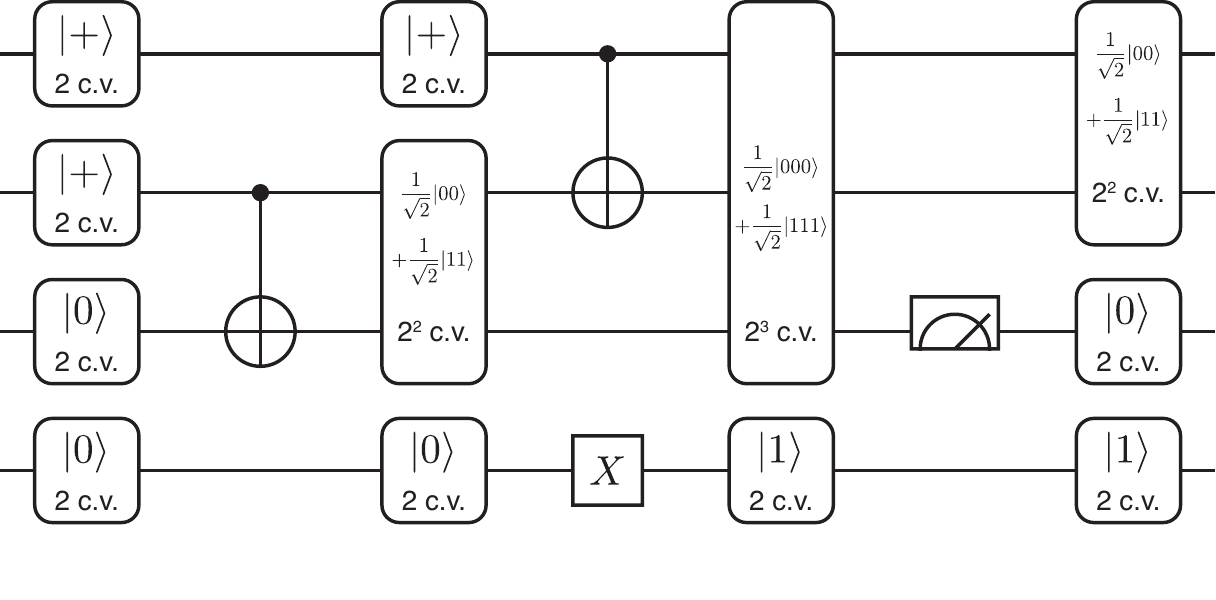}
    \caption{Simulating a deep lattice surgery computation is challenging, because there are constantly patches being entangled and measured out, but at a given time few are actually entangled. Each state is represented by certain number of complex variables (c.v.).}
    \label{fig:laztTensored}
\end{figure}

For verification purposes, we require a state-vector snapshot of the logical state of the lattice computation at every time-step. To meet both goals, we use a form of state-vector tracking that stores patch states without tensoring non-entangled states states together. This approach greatly extends what is possible to verify and debug compared to a naive state-vector simulator. Since this simulation is for debugging and verification purposes it is also able to detect when simulated states are numerically close enough to some common state such as $\ket{0}, \ket{+}, \ket{m}$ and display them as such. Figure~\ref{fig:laztTensored} is a graphical depiction of a circuit with intermediate states in a \emph{Lazily Tensored State-vector Simulation} (LTSvS).

\begin{figure}[!h]
     \centering
     \includegraphics[width=\columnwidth]{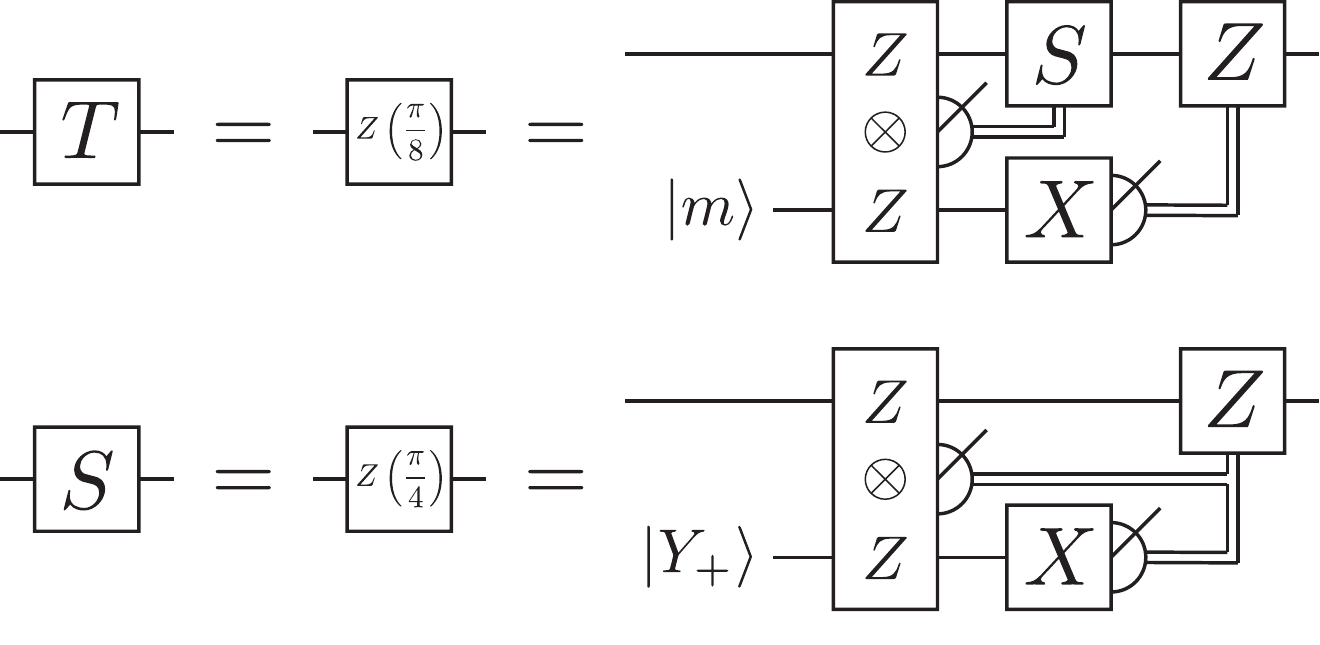}
     \caption{This figure only shows rotations by $\frac{\pi}{8}$ and $\frac{\pi}{4}$, but other angles with the same denominator are also possible by adjusting the classical logic controlling the corrective terms that follow. The $\frac{\pi}{8}$ Pauli rotations consume a distilled magic state, while the $\frac{\pi}{4}$ rotations consume a positive $Y$ eigenstate which is prepared by applying a twist based measurement\cite{litinski2018lattice}. }
     \label{fig:GatesToLLI}
\end{figure}

Conversion of gates and Pauli rotations to multi-body measurements and other LLI are presented in Fig.~\ref{fig:GatesToLLI}.

\begin{figure}[!h]
    \centering
    \begin{quantikz}
        & \ctrl{1} & \qw \\
        & \gate[wires=1]{{Z(\frac{\pi}{n})}} & \qw
        \end{quantikz}
        =\begin{quantikz}
        & \gate[wires=1]{{Z(\frac{\pi}{2n})}} & \ctrl{1} & \qw & \ctrl{1} &\qw \\
        & \gate[wires=1]{{Z(\frac{\pi}{2n})}} & \targ{} & \gate[wires=1]{{Z(-\frac{\pi}{2n})}} & \targ{}  & \qw
    \end{quantikz}
    \caption{Decomposing controlled rotations into CNOTs and single qubit rotations.}
    \label{fig:crz}
\end{figure}
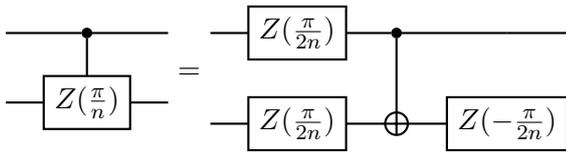

In Figure~\ref{fig:crz}, controlled rotations are common in circuit such as the QFT. The first step towards converting them to fault-tolerant instructions is breaking them down into single qubit rotations and CNOTs~\cite{kim2018gateidentity}. Single qubit rotations of angles greater than $\frac{\pi}{8}$ have to be approximated, while CNOTs we implement with lattice surgery~\cite{horsman2012surface}.

\begin{figure}[!h]
    \centering
    \includegraphics[width=\columnwidth]{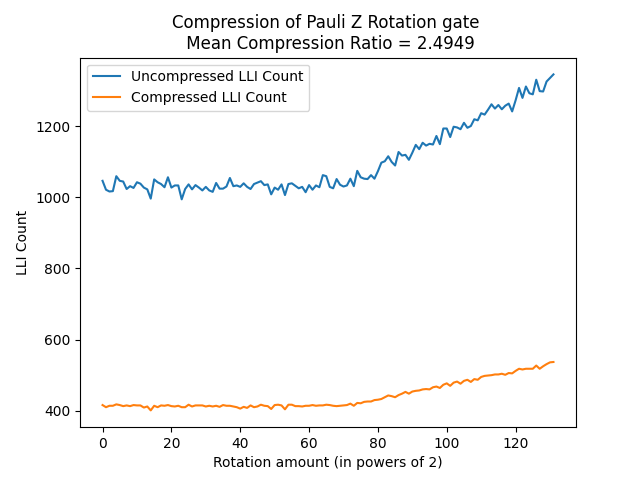}
    \caption{Benchmarking the decomposition techniques for arbitrary rotations to surface code instructions with and without Pauli rotation compression. The benchmark circuit is a single $R_Z(\frac{\pi}{2^n})$ rotation. It is possible to see how grouping gates that rotate in the same basis (as shown in Fig.~\ref{fig:gridsynthToRot}) drastically reduces the number of required surface code instructions.}
    \label{fig:pauli_compress}
\end{figure}

Fig.~\ref{fig:pauli_compress} illustrates the efficiency of the rotation compression technique we described in Fig.~\ref{fig:gridsynthToRot} from Sec.~\ref{sec:TransformingTheGateSet}. How much compression we get exactly depends on the types of gate sequences appearing in the approximations (e.g. both \code{HSSSTH} and a lone \code{T} compress to a single Pauli rotation). Testing on rotations from $R_Z(\frac{\pi}{2^{8}})$ to $R_Z(\frac{\pi}{2^{128}})$ we observed a factor of $\approx 2.5$ fewer LLI per small angle rotation -- this is an an illustration of our optimisation heuristics.

\subsection{Clifford Elimination}
\label{sec:ltalg}

According to the Gottesman-Knill theorem, it is possible to efficiently simulate circuits which only contain a particular set of gates, known as \textit{Clifford} gates~\cite{nielsen2001quantum,aaronson2004improved}. It is natural to ask whether it is possible to leverage classical computing to reduce the load on the QPU when processing such circuits. Litinski~\cite{litinski2019game} outlined an algorithm to remove the Clifford part of the circuit at compile time, when all we care about are measurement outcomes (i.e. we are not compiling the circuit to be a state preparation routine). We call this algorithm the Litinski Transform (LT) and provide an implementation in the following manner. 

The first step of LT is to convert each gate of the input circuit into a sequence of  rotations about $\frac{\pi}{2}$, $\frac{\pi}{4}$ or $\frac{\pi}{8}$, or multibody measurements with Pauli product observables. Of these blocks, only the $\frac{\pi}{8}$ rotations are not Clifford, so we apply Litinski's commutation rules to bring them all to the front of the circuit. Next the $\frac{\pi}{2}$ and $\frac{\pi}{4}$ rotations are commuted past the previously end-of-circuit measurements. Since this case we only care about measurement outcomes, the Clifford rotations that are now after measurement can be discarded.

\subsection{Layout File Format}

The layout determines how computation elements are placed on the lattice.  Herein we describe the technical details, limitations, examples and future enhancements of the layout format introduced in Sec.\ref{sec:routing}. The online source code repository includes more layout examples.

\textbf{Structure}. The purpose of our layouts is to define the structure of a lattice intended for lattice surgery operations, abstracting the details of the physical implementation. The layout is stored as an ASCII plain text file, in order to make it easy for humans to view and edit and to enable good portability (i.e. special tools needed to edit). The layout is specified by a grid of ASCII characters where each cell has a specific meaning based on the character it contains. For example, the text \code{QrQ} represents two logical qubits separated by an inactive routing space:

\includegraphics[width=0.5\columnwidth]{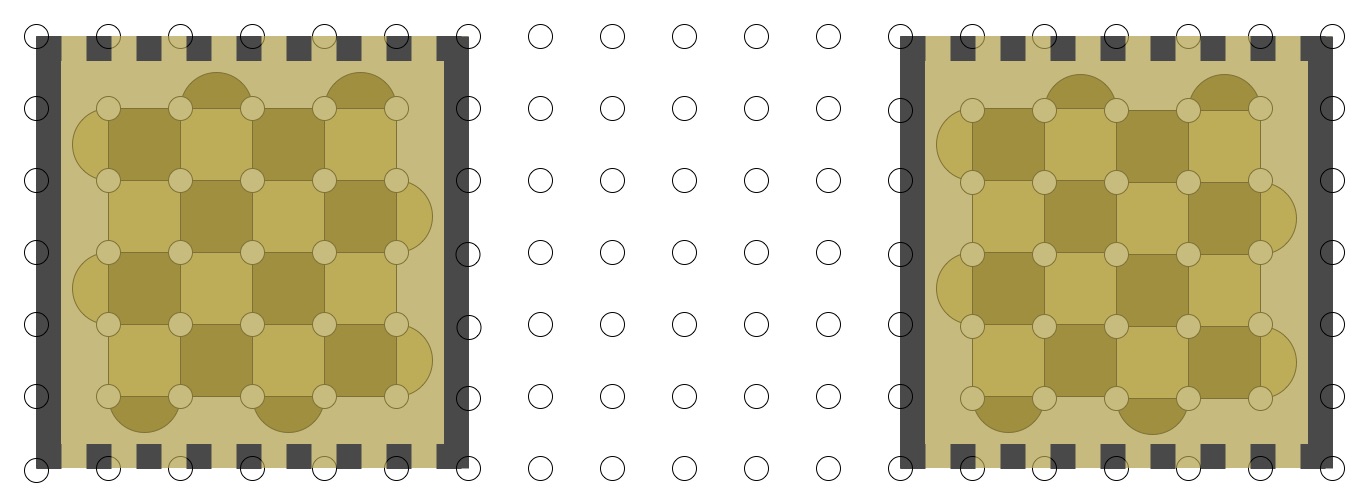}

The layout is always assumed to be rectangular. If a layout file is not rectangular in content, then the compiler will assume that the bounding box of it's contents is available and pad with empty routing space.

\textbf{Qubits} \code{Q}: Represents a patch holding a logical qubit encoded using the surface code planar code. The boundaries are assumed to be rough north-south, and smooth east-west, so it's the users responsibility that connectivity between each qubit is possible.

\textbf{Routing} \code{r}:  In their default or \emph{quiescent} state, these cells are inactive. This means they do not actively participate in quantum computations. However, when required, the contents of these routing cells can be "activated" to facilitate long-range merges and splits between distant \code{Q} cells, as shown in Fig.~\ref{fig:mergeSplit}.

\textbf{Ancilla} \code{A}: These cells are reserved to allocate new ancillae patches. Ancillae patches are auxiliary qubits used in quantum computations to assist in the construction of measurement-based gates. These patches can be in states like the $\ket{+}$ state, which mediates CNOT operations, or they can be places for the Y eigenstates used by $\frac{n}{4}$ Pauli rotations.

\textbf{Distillation regions} numbers \code{0-9}: Distillation regions are specialized areas  designated for the production of magic states, as defined in section \ref{sec:LogicalOps}. These regions are represented by areas with the same number in the ASCII layout. The extents of these regions are identified by running a connected components search on areas with matching numbers. This means that contiguous cells with the same number are considered part of the same distillation region.

The magic states produced by these regions are output to a neighboring \code{r} cell.  Distillation time is assumed to be the same for every distillation region. The compiler makes no assumptions about the internal operations or processes occurring within distillation regions. It is the user's responsibility to ensure that the size and configuration of a region is correct.

\textbf{Example 1}. The layout below supports two logical qubits, and routing. No two-qubit gates can be applied immediately, because there are no \code{A} patches.
\code{
\\
QrQ\\
rrr
}

\textbf{Example 2}. Seven qubits and four distillation regions can operate on the layout below. The \code{A} ancilla can be used for performing CNOTs between the logical qubits, for example.
\code{\\
rrrrrrr444\\
rQQrQQr444\\
rQQrQAr444\\
rrrrrrrrrr\\
r111222333\\
r111222333\\
r111222333
}

\textbf{Planned future extensions} of the layout file include: a) incorporation of multi-cell patches support; b) qubit indexing enhancements; c) initialization of square patches with alternate boundary configurations; d) introduction of a in browser editor equipped with syntax highlighting.

\clearpage

\end{document}